\documentclass[aps,prc,groupedaddress,showpacs,manuscript]{revtex4}
\usepackage{graphicx}

\begin{document}
\title {Probing the density dependence of the
symmetry potential with peripheral heavy-ion collisions}
\author {Yingxun Zhang$^{1}$
\email[]{zhyx@iris.ciae.ac.cn}, Zhuxia Li$^{1,2,3)}$
\email[]{lizwux@iris.ciae.ac.cn}}
\address{
 1) China Institute of Atomic Energy, P. O. Box 275 (18),
Beijing 102413, P. R. China\\
2) Center of Theoretical Nuclear Physics, National Laboratory of
Lanzhou Heavy Ion Accelerator,
 Lanzhou 730000, P. R. China\\
 3) Institute of Theoretical Physics, Chinese Academic of Science,
Beijing 100080}
\begin{abstract}
The peripheral heavy-ion collisions of $^{112, 124}Sn+ ^{86}Kr$ at
$E_{b}= 25AMeV$ are studied by means of the Improved Quantum
Molecular Dynamics Model(ImQMD). It is shown that the slope of the
average N/Z ratio of emitted nucleons vs impact parameters for
these reactions is very sensitive to the density dependence of the
symmetry energy. Our study also shows that the yields of $^{3}H$
and $^{3}He$ decrease with impact parameters and slope of the
yield of $^{3}H$ vs impact parameters as well as the ratio of
Y($^{3}H$)/Y($^{3}He$) depend on the symmetry potential strongly
for peripheral heavy-ion collisions.\\
\end{abstract}
 \maketitle

 Nowadays, the nuclear equation of state (EOS) for asymmetric
nuclear systems has attracted a lot of attention. The equation of
state for asymmetric nuclear matter can be described approximately
by the parabolic law
\begin{equation}
e(\rho,\delta)=e_{0}(\rho,0)+e_{sym}(\rho)\delta^{2},
\end{equation}
where $\delta=(\rho_{n}-\rho_{p})/(\rho_{n}+\rho_{p})$ is the
isospin asymmetry. $e_{0}$ is the energy per nucleon for symmetric
nuclear matter and $e_{sym}(\rho)$ is the bulk symmetry energy.
The nuclear symmetry energy term $e_{sym}(\rho)$ is very important
for understanding many interesting astrophysical phenomena, but it
is also a subject with large uncertainties, especially, its
density dependence\cite{Br00,BALi97,BALi02}. Therefor, to acquire
more accurate knowledge of the symmetry energy term becomes one of
the main goals in nuclear physics at present, and has driven a lot
of theoretical and experimental efforts. The facilities of rare
isotope beams provide opportunities to prepare and study the
dynamical evolution of nuclear systems with a range of isospin
asymmetries. This increases the domain over which a spatially
uniform local isospin asymmetry $\delta(r)$ may be achieved. The
isospin effect on the multifragmentation in central neutron-rich
heavy ion collisions at intermediate energies has been widely
studied both theoretically and experimentally and a amount of the
information for isospin dependence part, $e_{sym}(\rho)$, of
nuclear $EOS$ has been
obtained\cite{BALi97,Xu00,Dan02,BALi01,Li01,Tsa01,Tan,Ver01,Bar02}.
Recently, a large enhanced production of neutron rich rare
isotopes in peripheral collisions of Fermi energy for neutron rich
system was observed\cite{Sou03,Ves03}. For peripheral collisions,
the neck emission is the most important source. Under the
influence of the symmetry potential, the motion of neutrons and
protons towards to neck region will be different and thus the
ratio between the numbers of neutrons and protons in neck area
depends on the the density dependence of the symmetry potential
sensitively. Whether the neck emission in peripheral reactions
carries the information of the symmetry energy term? In this work
we investigate various observables in peripheral heavy ion
reactions at Fermi energies to seek those which are sensitive to
the symmetry potential. The calculations are performed for
peripheral reactions of $^{124}Sn+^{86}Kr$ and $^{112}Sn+^{86}Kr$
at incident energy of $25AMeV$.

For the theoretical description of heavy ion collisions, the
improved quantum molecular dynamic model (ImQMD)\cite{Wa02,Wa04}
is adopted in this work. The effective interaction potential
energy includes the nuclear local interaction potential energy and
Coulomb interaction potential energy,
\begin{equation}
U=U_{loc}+U_{Coul}.  \label{11}
\end{equation}
$U_{loc}$ is obtained from the integration of the nuclear local
interaction potential energy density functional. The nuclear local
interaction potential energy density functional
$V_{loc}(\rho(\mathbf{r}))$ in the ImQMD model reads
\begin{equation}
V_{loc}=\frac{\alpha }{2}\frac{\rho ^{2}}{\rho _{0}}+\frac{\beta }{\gamma +1}%
\frac{\rho ^{\gamma +1}}{\rho _{0}^{\gamma }}+\frac{g_{sur}}{2\rho _{0}}%
(\nabla \rho )^{2}+\frac{C_{s}}{2\rho _{0}}(\rho ^{2}-\kappa
_{s}(\nabla \rho )^{2})\delta ^{2}. \label{13}
\end{equation}
The first three terms in above expression can be obtained from the
potential energy functional of Skyrme force directly. The fourth
term is the symmetry potential energy where both the bulk and the
surface symmetry potential energy are included. The surface
symmetry potential energy term modifies the symmetry potential at
the surface region and it is important for having a correct
neutron skin and neck dynamics in heavy ion collisions. The
Coulomb energy includes both the direct and exchange
contributions. In the collision term, isospin dependent
nucleon-nucleon scattering cross sections are used\cite{Cug96} and
the Pauli blocking effect is treated as in\cite{Li01}. We have
applied this model to study heavy ion reactions at intermediate
energies. The relevant observable to present study is the charge
distribution in heavy ion collisions at intermediate energies.  As
an example, in Fig.1 we show the charge distribution of products
in the central collisions of $^{129}Xe+^{124}Sn$ at 50 AMeV and
$^{40}Ca+^{40}Ca$ at 35 AMeV.  The experimental data from
\cite{Hud03,Hag94} are also given in the figure. Clusters are
constructed by means of the coalescence model widely used in QMD
calculations in which particles with relative momenta smaller than
$P_{0}$ and relative distances smaller than $R_{0}$ are coalesced
into one cluster(here $R_{0}=3.0fm$ and $P_{0}=250MeV$/c) is
adopted). Concerning light charged particles of $Z \leq 2$
emission, we further introduce the mechanism of the coalescence of
free nucleons adopted by Neubert and Botvina(mec of N.B.) in ref.
\cite{Neu00} where this mechanism was shown to be important for
light charged particles. We find that the charge distribution at
$Z \leq 2$ region is better described by introducing the
production of light charged particles by the coalescence of free
nucleons adopted in ref. \cite{Neu00}. One can see that the
calculation results are in good agreement with experiments. Now we
apply this model to seek the sensitive observable for testing the
density dependence of the symmetry potential in heavy ion
collisions at energies around Fermi energy. In Eq.3 a linear form
of the density dependence of symmetry potential is used. In order
to study the dynamical effect of the density dependence of the
symmetry potential on heavy ion collisions by means of the
transport theory, we verify the form of the density dependence of
the bulk symmetry potential energy in the potential energy density
functional. For simplicity, the bulk symmetry potential energy
density is taken to be a form of
\begin{equation}
w_{sym}(\rho)=\frac{C_{s}}{2}u^{\gamma}\rho\delta^{2},
\end{equation}
where $u=\rho/\rho_{0}$. In the following, we refer $\gamma=0.5$
to soft symmetry potential ($soft-sym$) case and $\gamma=1.0$ to
the stiff symmetry potential ($stiff-sym$) case. The single
particle symmetry potential can be derived from Eq.(4) and reads
\begin{equation}
v^{q}_{sym}=\frac{\partial w_{asy}}{\partial
\rho_{q}}=\frac{C_{s}}{2}[(\gamma-1)u^{\gamma}\delta^{2}\pm2u^{\gamma}\delta]
\end{equation}
where $q$ denotes neutron and proton and symbols "+" and "-" are
for neutron and proton, respectively. The parameters used in this
work are listed in the Table.1.

\begin{table}[htbp]
 \caption{The parameter adopted in the work} \label{Table.1}
\begin{tabular}{ccccccc}
\hline
\hline
  $\alpha(MeV)$ & $\beta(MeV)$ & $\sigma$ & $g_{sur}(MeVfm^{2})$ & $\kappa_{s}(fm^{2})$ & $C_{s}(MeV)$ & $\gamma$ \\
\hline
  $-356$ & $303$ & $7/6$ & $22$ & $0.08$ & $32$ & $0.5or1.0$ \\
\hline
\hline
\end{tabular}
\end{table}

Fig.2 shows (a)the symmetry energy term $e_{sym}(\rho)$ in the
asymmetry nuclear matter for the $soft-sym$ case and the
$stiff-sym$ case, (b)the single particle symmetry potential
$v^{n(p)}_{sym}$, and (c)the chemical potential for neutrons and
protons. The figure shows that the single particle symmetry
potential for neutron and proton as well as the chemical potential
of neutron and proton depend on the form of the density dependence
of the symmetry potential strongly, which will influence the
motion of neutrons and protons, especially, the motion towards to
neck. We will see this effect on the N/Z ratio at neck region
later. The difference between the neutron and proton chemical
potentials is given by
$\mu_{n}-\mu_{p}=4(\frac{C_{s}}{2}u^{\gamma}+\frac{1}{3}\epsilon^{0}_{F}u^{2/3})=4e_{sym}(\rho)\delta$.
, which is directly related with the emission rate of neutrons and
protons.

To obtain the general information of the process of the peripheral
reactions, we show the time evolution of contour plots of neutron
and proton densities for a typical event of reaction
$^{124}Sn+^{86}Kr$ at $E_{b}=25AMeV$, $b=10fm$ within the time
period of $25fm/c$ to $200fm/c$ in Fig.3. Calculations are
performed by using $stiff-sym$. From Fig.3 one can see that the
neutrons reach the neck region earlier than the protons. After the
touching of projectile and target at about $50fm/c$, there is a
very gentle compression in the neck region. At this stage the
density at the neck region reaches a value greater than the normal
density. And after then, this region expands and finally ruptures
during $125fm/c-200fm/c$. While, the density of $PLF^{*}/TLF^{*}$
is always around the normal density and it means that there is no
compression happening for $PLF^{*}/TLF^{*}$. Thus, nucleons, light
charged particles and very few intermediate mass fragments, are
emitted mainly in the neck region . In Fig.4, we show the time
evolution of the average $N/Z$ ratio of the nucleons in a cube
with a side length of $8fm$ centered at the point where it has an
equal distance from the surface of projectile and target(see the
small inserted figure in Fig.4(a)) for reactions of
$^{112,124}Sn+^{86}Kr$. The left panel is the results with
stiff-symmetry potential and the right one is those with
soft-symmetry potential. The general behavior of the time
evolution of the N/Z ratio of nucleons at neck region is: At about
25 fm/c it reaches a very high value because neutrons move faster
than protons at early time, then it reduces till it reaches a
minimum at about 100fm/c-110fm/c depending on the symmetry
potential being stiff or soft, and then remains at this state for
a while which form a plateau. The N/Z value of nucleons at the
plateau is higher for the $soft-sym$ case than for the $stiff-sym$
case for the same reactions. After the plateau, it gradually
increases. The increasing slope depends on the impact parameters,
isospin asymmetry of the system and the symmetry potential. The
larger the asymmetry of the system and the stiffer of the symmetry
potential is, the larger the increasing slope is at the fixed
impact parameters. The behavior of the time evolution of the N/Z
ratio at neck area shown in Fig.4 clearly reveals the effects from
both the neutron skin and the symmetry potential. As is shown in
Fig.2 that the neutron chemical potential from high density to the
minimum for the $stiff-sym$ case decreases more steeply than that
for the $soft-sym$ case and vice versa for proton chemical
potential, which leads to the enhancement of both the N/Z ratio
and the increasing slope of the N/Z ratio for the $stiff-sym$ case
compared with that for the $soft-sym$ case. While the increase of
the N/Z ratio with impact parameters reveals the neutron skin
effect in peripheral heavy ion collisions. The comparison between
reactions of $^{124}Sn+^{86}Kr$ and $^{112}Sn+^{86}Kr$ indicates
much stronger impact dependence for the former reaction and it can
be attributed to the stronger neutron skin effect for the former
case. The different N/Z ratio of nucleons in the neck area with
the soft and stiff symmetry potential will lead to different
ratios between the neutron to proton emission rate and the
isobaric ratio of yields of $^{3}H $ and $^{3}He$, and other
particles, which emit from the neck area.

How does the emission rates of neutrons and protons from the neck
area depend on the density dependence of the symmetry potential?
Fig.5 shows the average $N/Z$ ratio of emitted free nucleons and
heavy residues ($Z\ge30$) versus impact parameters. The left panel
shows the results for $^{112}Sn+^{86}Kr$ and the right panel for
$^{124}Sn+^{86}Kr$. From the figure one sees that the $N/Z$ ratio
of emitted nucleons is higher with soft symmetry potential than
with stiff symmetry potential. This behavior has been explored in
\cite{BALi97}. As far as the impact parameter dependence is
concerned, the N/Z ratio is rather flat for central collisions(at
$b=2,4fm$), while it increases with the impact parameters for
peripheral collisions($b=7-10fm$).  The increasing slope of the
$N/Z$ ratio of emitted nucleons with impact parameters is
sensitive to the symmetry potential. For reactions of
$^{112}Sn+^{86}Kr$ and $^{124}Sn+^{86}Kr$, the increasing slope of
the $N/Z$ ratio of emitted free nucleons versus impact parameters
calculated with $stiff-sym$ is about 2-3 times larger than that
with $soft-sym$ in the range of $b=7fm$ - $10fm$. And the slope
for $^{124}Sn+^{86}Kr$ is larger than that for $^{112}Sn+^{86}Kr$.
The behavior of the dependence of the N/Z ratio of emitted
nucleons on impact parameters, the isospin asymmetry of systems
and symmetry potentials shown in Fig.5 is closely related with the
different behaviors of the time evolution of N/Z ratio of nucleons
in neck area at corresponding conditions shown in Fig.4.
Correspondingly, the $N/Z$ ratio of heavy residues is larger for
$stiff-sym$ case than for $soft-sym$ case due to the balance of
emitted protons and neutrons
 and those in heavy residue. The detailed analysis for heavy
 residues is in progress. For the same reason, the slope of the $N/Z$
ratio of heavy residues vs impact parameters also depends on the
symmetry potential. We find that introducing the production of
light charged particles by the coalescence of free nucleons does
not change the feature of the influence of the different forms of
the symmetry potential on the N/Z ratio of emitted nucleons vs
impact parameters.

Fig.6 shows the yields of $^{3}H$ and $^{3}He$  as a function of
impact parameters for $^{124}Sn+^{86}Kr$ (left panel) and
$^{112}Sn+^{86}Kr$ (right panel) at $25AMeV$ calculated with
$stiff-sym$ and $soft-sym$ . The yield of $^{3}H$ is greater than
that of $^{3}He$ for the same reaction system calculated with the
same symmetry potential for all impact parameters. For central
collisions( at $b=2,4fm$), the ratio between the yields of $^{3}H$
and $^{3}He$ for $^{124}Sn+^{86}Kr$ is about 2.5 with $soft-sym$
and 1.9 with $stiff-sym$. For reactions of $^{112}Sn+^{86}Kr$, it
is about 1.98 with $soft-sym$ and about 1.54 with $stiff-sym$.
Introducing the production of light charged particles by
coalescence of free nucleons adopted in \cite{Neu00} largely
increases the yields of $^{3}H$ and $^{3}He$ but does not change
the behavior of the dependence on impact parameters. The ratio
between yields of $^{3}H$ and $^{3}He$ at central collisions
changes slightly. They are about 2.33 with $soft-sym$ and 1.86
with $stiff-sym$ for reactions of $^{124}Sn+^{86}Kr$ and about
1.85 with $soft-sym$ and about 1.44 with $stiff-sym$ for
$^{112}Sn+^{86}Kr$. These values are roughly coincident with the
experiment results of ref. \cite{Lef99} where the yields of
$^{3}H$ and $^{3}He$ for central collisions of $^{36}Ar+^{58}Ni$
at $Eb=74AMeV$ were measured and the ratio between them was about
1.4. For peripheral collisions($b=7-10fm$), the yields of $^{3}H$
and $^{3}He$ decrease with impact parameters owing to the
decreasing of the size of the neck area with impact parameters.
While the decreasing slope depends on the symmetry potential
adopted, especially for the yield of $^{3}H$. The average
decreasing slope of the yield of $^{3}H$ vs impact parameters for
reactions $^{112}Sn+^{86}Kr$ and $^{124}Sn+^{86}Kr$ calculated
with $soft-sym$ is about $1.47$ and $1.59$ times larger than those
with $stiff-sym$, respectively. The average decreasing slope of
the yield of $^{3}He$ vs impact parameters is less sensitive to
the symmetry potential.

In summary, we have investigated the dynamical effects of the
symmetry energy on the peripheral heavy ion collisions for
$^{124,112}Sn+^{86}Kr$ at $25AMeV$ by means of the $ImQMD$ model.
Our investigation shows that the neck emission is the main
emission source of nucleons and light charged particles for
peripheral reactions. The N/Z ratio at neck area strongly depends
on the impact parameter, the neutron excess of the system and the
the stiffness of the symmetry potential which leads to the N/Z
ratio of emitted nucleons and the light charged particles changing
with impact parameters, systems and the stiffness of the symmetry
potentials for peripheral reactions. Our results show that the
average $N/Z$ ratio of emitted nucleons calculated with soft
symmetry potential is higher than that with stiff symmetry
potential. Furthermore, we have shown that the N/Z ratio increases
with impact parameters and the increasing slope with stiff
symmetry potential is about twice as large as those with soft
symmetry potential. The average $N/Z$ ratio of heavy residues
shows the opposite trend but much weaker.  Our results show that
the yields of $^{3}H$ and $^{3}He$ decrease with impact parameters
for peripheral reactions and the ratio between the yields of
$^{3}H$ and $^{3}He$ depends on the symmetry potential. And the
reducing slope of the yield of $^{3}H$ with impact parameters also
depends on the symmetry potential strongly. Our calculations also
show that introducing the production of light charged particles by
the coalescence of free nucleons increases the yields of $^{3}H$
and $^{3}He$ but does not change the feature of the influence of
different stiffness of the symmetry potential on the N/Z ratio of
free nucleons, the yield of $^{3}H$ and the ratio between the
yields of $^{3}H$ and $^{3}He$ and their dependence on impact
parameters. From our study we can conclude that for peripheral
neutron-rich heavy ion collisions, the slope of average N/Z ratio
of emitted nucleons, the yield of $^{3}H$  and the ratio between
the yields of $^{3}H$ and $^{3}He$ with respect to impact
parameters are sensitive to the density dependence of the symmetry
potential and can be used to prob the density dependence of the
symmetry potential.

\begin{center}
{\bf ACKNOWLEDGMENTS}
\end{center}
 This work is supported by National Natural
Science Foundation of China under Grant Nos.10175093,
10235030,10235020 and by Major State Basic Research Development
Program under Contract No.G20000774.

\newpage
\begin{description}
\item[\texttt{Fig.1}] The charge distribution of products in the
central collisions for reactions $^{129}Xe+^{124}Sn$ at 50 AMeV
and $^{40}Ca+^{40}Ca$ at 35 AMeV. The open circles and squares are
the results with and without the productions of light charges
particle by the coalescence of free nucleons adopted by
\cite{Neu00}. The triangles are the experimental data taken from
\cite{Hud03} for $^{129}Xe+^{124}Sn$ and \cite{Hag94} for
$^{40}Ca+^{40}Ca$.
 \item[\texttt{Fig.2}] (a)The symmetry energy, (b)the single
particle symmetry potential and (c)the chemical potential for
proton and neutron in the nuclear matter at $\delta=0.1,0.3$
calculated with the soft symmetry potential and stiff symmetry
potential, respectively.

\item[\texttt{Fig.3}] The time evolution of contour plots of
proton and neutron densities for reactions of $^{124}Sn+^{86}Kr$
at $Eb=25AMeV$ and $b=10fm$, calculated with the stiff symmetry
potential.

\item[\texttt{Fig.4}] The time evolution of the average $N/Z$
ratio for nucleons in the neck region (see the text) for reactions
of $^{124}Sn+^{86}Kr$ and $^{112}Sn+^{86}Kr$ at impact parameters
of $b=7-10$. The calculations are performed with $stiff-sym$  and
$soft-sym$, respectively.

\item[\texttt{Fig.5}] The average $N/Z$ ratios of the emitted free
nucleons(circles) and heavy residues with $Z\geq30$(squares) vs
impact parameters with $soft-sym$ (open symbols) and $stiff-sym$
case(solid symbols), respectively. The Left panel is for
$^{112}Sn+^{86}Kr$ and the right panel for $^{124}Sn+^{86}Kr$.

\item[\texttt{Fig.6}] The yields of $^{3}H$(squares) and
$^{3}He$(circles) vs impact parameters for reactions of
$^{124}Sn+^{86}Kr$(right) and $^{112}Sn+^{86}Kr$(left) with
$soft-sym$(open symbols) and $stiff-sym$(solid
symbols),respectively. The solid and dashed lines are the results
with and without the production of light charged particles by the
coalescence of free nucleons, respectively.

\end{description}

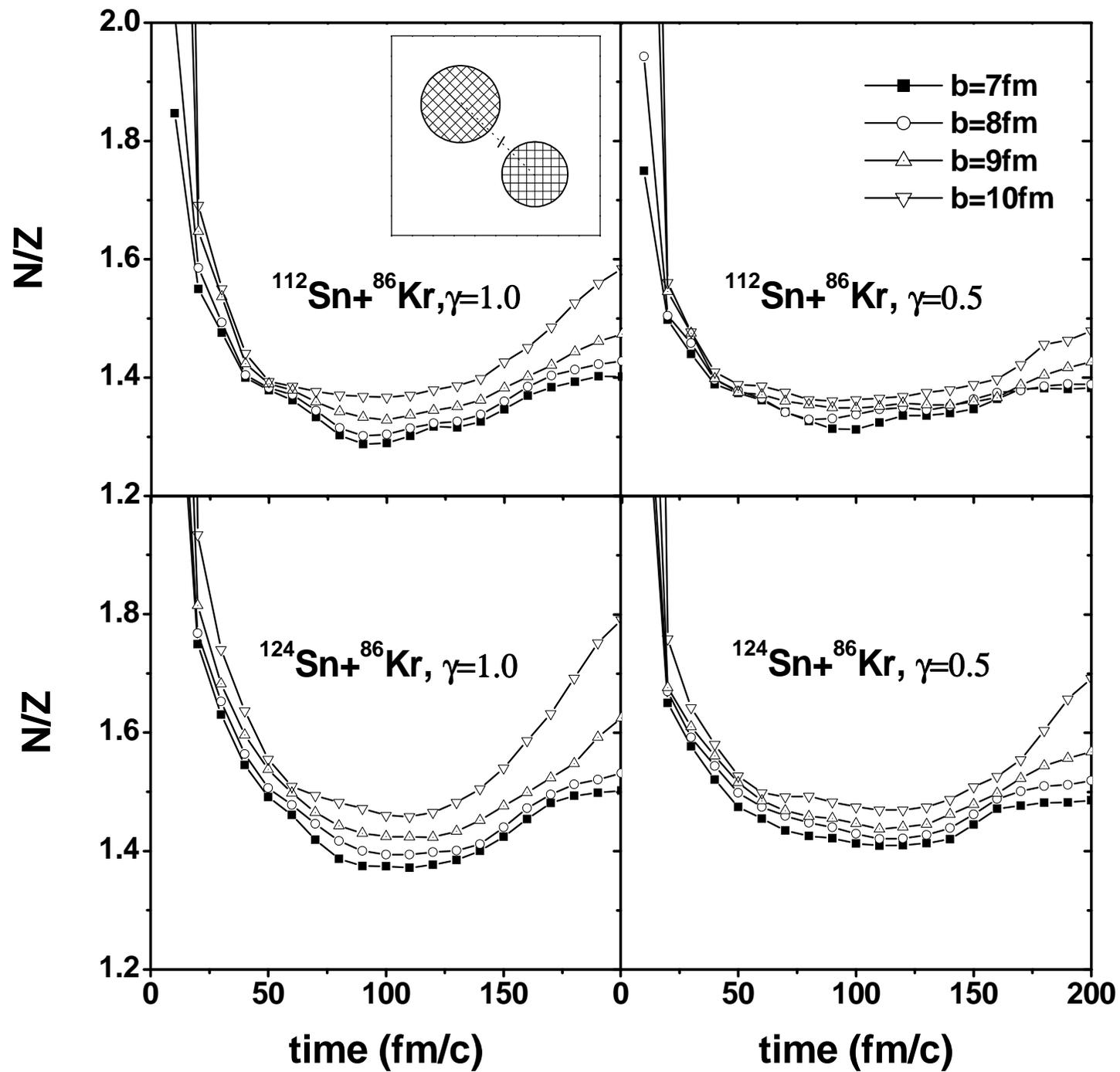
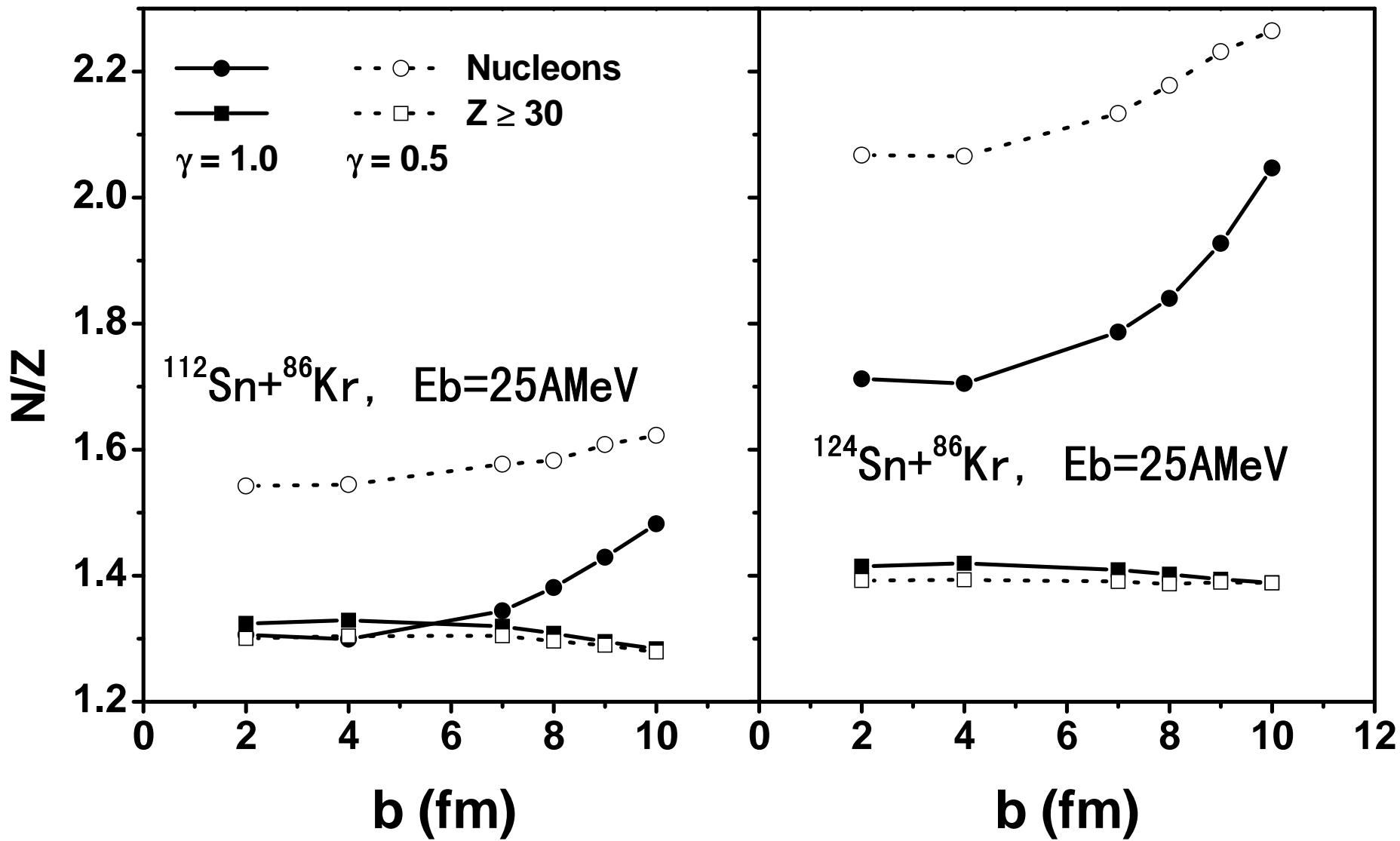
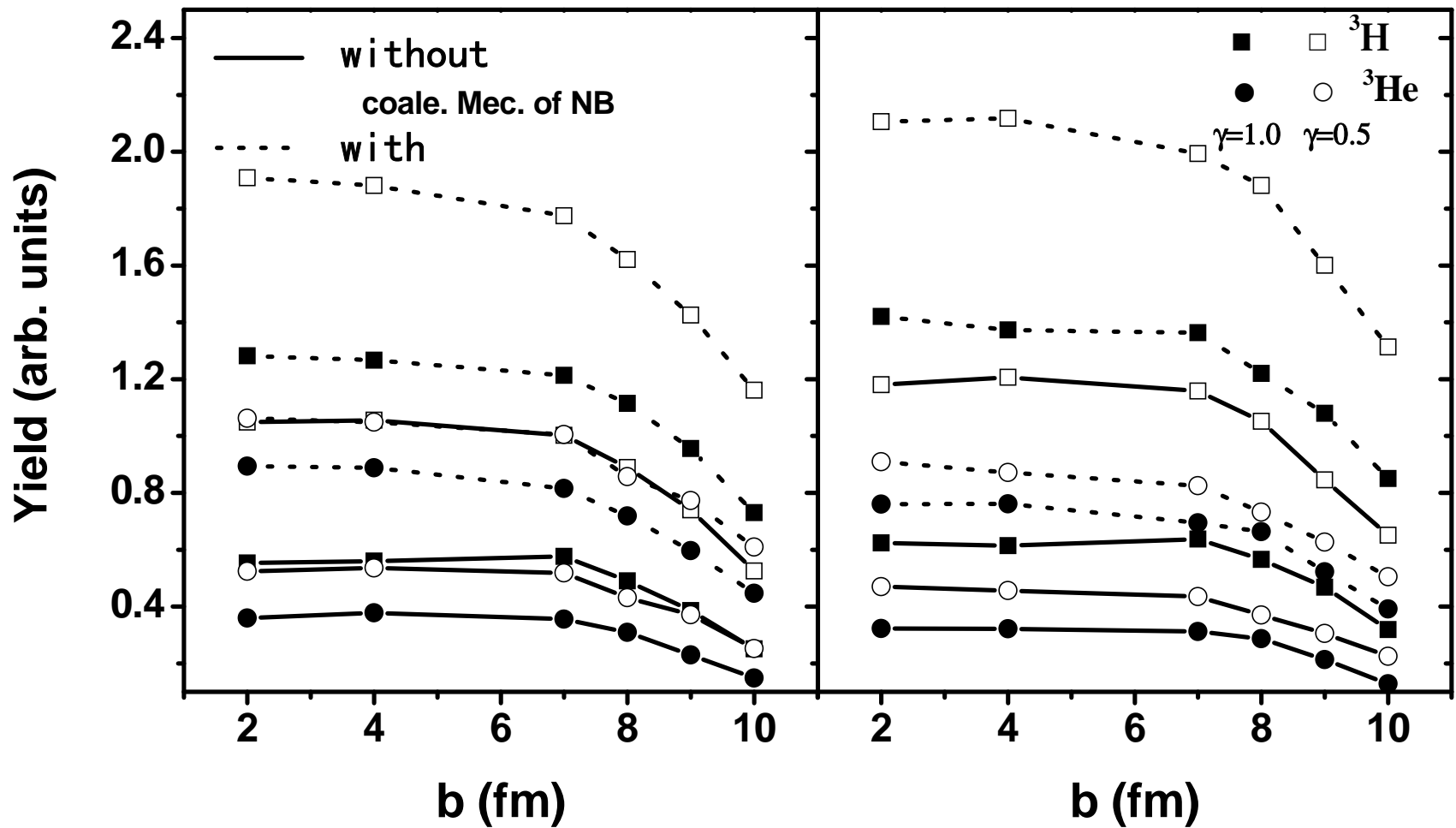
\end{document}